# The cross sections for inelastic neutron scattering
# and
# radiative capture in the field of a light wave

E.A. Ayryan [1], A.H. Gevorgyan [2], I.V. Dovgan [3], K.B. Oganesyan [1,4] *

*bsk@yerphi.am

[1] Joint Institute for Nuclear Research, LIT, Dubna, Moscow Region

[2] Yerevan State University, Yerevan, Armenia

[3] Department of Physics, Moscow State Pedagogical University, Moscow 119992, Russia

[4] A.I. Alikhanyan National Science Lab, Yerevan Physics Institute, Yerevan, Armenia

The mixing of the levels of a compound nucleus in the field of a high-intensity light wave is considered. The cross sections for inelastic neutron scattering and radiative capture are computed with allowance for this effect. The effect of the electron shell on the effective charge for the dipole interaction between the neutron + nucleus system and the external electromagnetic field is taken into consideration.

## 1. INTRODUCTION

The effect of resonance excitation of atoms in a laser radiation field has been fairly well studied from both the theoretical and experimental points of view [1]. In the case of a nucleus the resonance excitation of the levels near the ground state was accomplished with the aid of the Mossbauer effect. There are only incoherent-radiation sources for this energy region. The situation is different when the levels of a compound nucleus are excited, e.g., in reactions with neutrons. In the case of heavy nuclei (the rare earths, the actinides), for excitations corresponding to the binding energy of the neutron, the distance between the levels of the compound nucleus is close in order of magnitude to the energies of laser-light quanta. Therefore, we can, in principle, raise the question of resonance transitions between two levels of a compound nucleus, one of which is excited in the course of the neutron capture. In this case levels with opposite parities, which may differ significantly in the partial nuclear decay probabilities associated with them, will, as a rule, be mixed in an external resonance electromagnetic field.

In earlier papers [2,3] was attempted to estimate the probability for mixing of the levels of a compound nucleus in the presence of an external electromagnetic field. There was assumed that the electron shell of the atoms had no effect on the probability for resonance mixing of the levels of a compound nucleus, i.e., that the frequency of the electromagnetic field was not equal to the electron-transition frequency. Furthermore, the spin of the neutron was assumed to be equal to zero. These two assumptions are rather far removed from what the real situation is, especially, the neglect of the effect of the electron shell. It had been shown earlier in another connection [4] that, in the case of electric dipole nuclear transitions with low frequency ($\omega_{nucl} \leq \omega_{el}$ the electron - transition frequency) ), the electron shell of the atom has an appreciable effect on the nuclear - transition probabilities.

In the present paper we investigate the role of the electron shell in the processes of resonance mixing of the levels of a compound nucleus, and also compute the probability for this mixing.

## 2. EFFECT OF THE ELECTRON SHELL

The Hamiltonian of an atom + neutron system in an external electromagnetic field has the form

$$H = \frac{1}{2m}\sum_{i=1}^{z'}\left(\mathbf{p}_i + \frac{e}{c}\mathbf{A}\right)^2 + \sum_{i=1}^{z'}V(|\mathbf{r}_i - \mathbf{r}_{nucl}|) + \frac{\mathbf{p}_n^2}{2M} + V(|\mathbf{r}_n - \mathbf{r}_{nucl}|) + \frac{1}{2MA}\left(\mathbf{p}_{nucl} - \frac{Ze}{c}\mathbf{A}\right)^2 + H_{\text{int}} \quad (1)$$

where m and M are respectively the electron and neutron masses, $\mathbf{p}_i$ is the electron momentum operator, $\mathbf{p}_n$ is the incident-neutron momentum operator, $\mathbf{p}_{nucl}$ is the nucleus momentum operator, MA is the mass of the nucleus, $V(|\mathbf{r}_n - \mathbf{r}_{nucl}|)$ is the effective neutron-nucleus interaction potential, $V(|\mathbf{r}_i - \mathbf{r}_{nucl}|)$ is the potential energy operator of the atom, and $H_{\text{int}}$ is the operator for the interaction of the neutron with the nucleons of the nucleus. The $H_{\text{int}}$ interaction leads to the excitation of more complicated configurations, e.g., three-quasiparticle configurations.

Let us write the vector potential of the external electromagnetic field in the dipole approximation in the form

$$\mathbf{A} = \mathbf{A}_0 \cos \omega t. \qquad (2)$$

We transform to a system of coordinates fixed to the nucleus; to do this, we introduce the relative coordinates and the coordinate of the center of gravity:

$$\boldsymbol{\xi}_i = \mathbf{r}_i - \mathbf{r}_{nucl}, \quad \boldsymbol{\xi}_n = \mathbf{r}_n - \mathbf{r}_{nucl},$$
$$\mathbf{X} = \left( M\mathbf{r}_n + AM\mathbf{r}_{nucl} + m\sum_{i=1}^{z'} \mathbf{r}_i \right) / (Z'm + M(A+1)). \qquad (3)$$

After simple transformations, the Hamiltonian can be written in the form

$$H = H_{oa} + H_{on} + V, \qquad (4)$$

where $H_{oa}$ and $H_{on}$ are respectively the Hamiltonians of the atom and neutron + nucleus system in the absence of an external electromagnetic field, and for the operator V we obtain the following expression:

$$V = -i\hbar \frac{e}{mc} \sum_{i=1}^{z'} \frac{\partial}{\partial \xi_i} \mathbf{A} - i\hbar \frac{Ze}{MA} \frac{\partial}{\partial \xi_n} \mathbf{A} - \frac{\hbar^2}{MA} \sum_{i=1}^{z'} \frac{\partial^2}{\partial \xi_i \partial \xi_n} \mathbf{A}. \qquad (5)$$

Here the first term describes the interaction of the electrons with the external electromagnetic field, the summation being over all the electrons in the atom ($Z' \leq Z$); the second term describes the interaction of the system neutron + nucleus with the A field; and the third term describes the correlation between the motion of the electron + nucleus system and that of the neutron. This interaction is the result of the separation of the center of mass of the atom + neutron system and the going over to the coordinates of the neutron and the electrons relative to the nucleus. In [4] a similar interaction between the nucleons of the nucleus and the electron shell arises from the Coulomb energy as a result of an expansion in powers of the small parameter $|\mathbf{r}_{nucl}|/|\mathbf{r}_{el}|$ Therefore, this interaction is proportional to the square of the charge, but the interaction considered in the present paper does not depend on the charge at all.

The quantity Ze/A should be interpreted as the effective charge for the electric dipole transitions in the neutron- + nucleus system. If the effect of the electron shell is insignificant, then only this charge needs to be considered. Let us consider the effect of the electron shell on the effective charge. To do this, let us compute the matrix element of the transition between the compound-nucleus level for the s neutron and the corresponding p level with emission or absorption of a quantum of the external electromagnetic field. We shall assume that the electron shell of the atom remains in the process in the ground state. Then we obtain in the lowest orders of perturbation theory the expression

$$V_{\lambda n} = \frac{Ze}{2MAc}(\mathbf{p}_n \mathbf{A}_0)_{\lambda n} - \frac{e}{2mMAc}\sum_m \left(\sum_{i=1}^{z'}\mathbf{p}_i\mathbf{p}_n\right)_{\lambda n, 0m}$$
$$\times \left(\sum_{i=1}^{z'}\mathbf{p}_i\mathbf{A}_0\right)(E_m - \hbar\omega - E_0)^{-1} - \frac{e}{2mMAc}\sum_m \left(\sum_{i=1}^{z'}\mathbf{p}_i\mathbf{A}_0\right)_{0m}\left(\sum_{i=1}^{z'}\mathbf{p}_i\mathbf{A}_0\right)(E_m + E_\lambda - E_0 - E_n)^{-1},$$
(6)

where $\mathbf{p}_i = -i\hbar\partial/\partial\xi_i$, $\mathbf{p}_n = -i\hbar\partial/\partial\xi_n$, the sum is over all the excited electron states of the atom, $E_m$ and $E_0$ are respectively the excited- and ground-state energies of the atom, $E_n$ is the energy of the incoming neutron, and $E_\lambda$ is the energy of the compound-nucleus level at which the neutron is captured. If the absorption of the neutron is accompanied by the absorption of a photon, then $E_\lambda = E_n + \hbar\omega$. Substituting this value of $E_\lambda$ into (6), and using the relation

$$\mathbf{p}_{12} = im(E_1 - E_2)\mathbf{r}_{12}/\hbar,$$

we obtain ($A_0 = cE_0/\omega$)

$$V_{\lambda n} = i\frac{Ze}{2A}(\xi_{nz}E_0)_{\lambda n} - i\frac{em}{2\hbar A}(\xi_{nz}E_0)_{\lambda n}\left(\sum_m \frac{\omega_{m0}^2}{\omega_{m0} - \omega}|(\xi_z)_{m0}|^2 + \frac{\omega_{m0}^2}{\omega_{m0} + \omega}|(\xi_z)_{m0}|^2\right).$$
(7)

Here, for simplicity, we consider the case in which the z axis is oriented along the vector $\mathbf{E}_0$. Let us transform the expression in the round brackets. Let us separate out in this expression the $\omega$-independent term, and, using the well-known sum rule for the square of the matrix element of the dipole operator [5] reduce it to the ratio Z'/Z, where Z' is the number of electrons in the atom (ion) and Z is the nuclear charge. Let us express the other term, which is proportional to $\omega^2$, in terms of the dipolar polarizability of the atom (ion) [1]:

$$\beta(\omega) = \frac{e^2}{\hbar}\sum_m |(\xi_z)_{m0}|^2\left(\frac{1}{\omega_{m0} - \omega} + \frac{1}{\omega_{m0} + \omega}\right).$$
(8)

As a result, we obtain an expression for the effective transition matrix element for the neutron + nucleus system in the form

$$V_{\lambda n} = i\frac{Ze}{2A}(\xi_{nz}E_0)_{\lambda n}\left(\frac{Z - Z'}{Z} - \frac{m\omega^2}{\hbar c}\frac{137}{Z}\beta(\omega)\right) = i\frac{Ze}{2A}(\xi_{nz}E_0)_{\lambda n}\Delta.$$
(9)

Estimates of the second term in (9) in the static limit yield a value of $10^{-4}$ for non alkali neutral atoms and a value roughly an order of magnitude higher for the alkali atoms. When allowance is made for the $\omega$ dependence of the dynamical polarizability, the corresponding term can be much greater than the term in the static case. Thus, for example, for the cesium atom and a neodymium laser the value of the polarization term can be as high as $10^{-2}$. In the case of ions the first term, as a rule, predominates. Thus, the electron shell has quite an appreciable effect on the effective charge for the electric dipole transitions in the neutron + nucleus system in a laser-radiation field. Notice that this conclusion is valid in the nonresonance-with respect to the electrons approximation (i.e., in the case when $\omega_{m0} - \omega \gg \Gamma_m$), where $\Gamma_m$ is the width of the corresponding atomic level.

## 3. PROBABILITY FOR MIXING OF THE LEVELS OF A COMPOUND NUCLEUS

For the subsequent calculations it is convenient to express the nuclear transition operator in (9) in terms of the spherical harmonics $Y_{lm}(\theta, \varphi)$:

$$\hat{V} = e_t E_0 \xi_n \sqrt{\frac{\pi}{3}} \left[ \cos\theta_0 Y_{10} + \frac{1}{\sqrt{2}} \sin\theta_0 (Y_{i-1} \exp(i\varphi_0) - Y_{i1} \exp(-i\varphi_0)) \right] \quad (10)$$

where $e_t = Ze/A$, $\theta_0$ is the angle between $\mathbf{E}_0$ and $\mathbf{p}_n$, $\varphi_0$ is the corresponding azimuthal angle, and $\theta$ is the angle between $\mathbf{p}_n$ and $\xi_n$.

We shall compute the cross section for an inelastic scattering in which the neutron undergoes a transition from the state with initial energy $\varepsilon_p$ into the state with energy $\varepsilon_p \pm \hbar\omega$, using perturbation theory. Then the cross section for inelastic scattering of the neutron in a laser-radiation field can be written in the form

$$d\sigma_{\lambda n} = \frac{2\pi M}{\hbar^2 k_0} \sum_i |V_{\lambda n}|^2 \delta(\varepsilon_p \pm \hbar\omega - E_1), \quad (11)$$

where $k_0^2 = 2M\varepsilon_p / \hbar^2$ is the square of the wave vector of the incident neutrons; the summation is over all the final states. In the case of slow-neutron scattering by nuclei we can limit ourselves in the initial state to the consideration of the s and p waves. We shall assume that the target nucleus has zero spin; then the initial-state wave function in the region $r_n > R$ (the nuclear radius) will have the form (we assume, for simplicity, that the neutron is a spinless particle)

$$\varphi_{init} = \exp(i\delta_0(k_0))[\cos\delta_0(k_0)j_0(x_0) - \sin\delta_0(k_0)\eta_0(x_0)]P_0(\cos\theta)$$
$$+3i\exp(i\delta_1(k_0))[\cos\delta_1(k_0)j_1(x_0) - \sin\delta_1(k_0)\eta_1(x_0)]P_1(\cos\theta), \quad (12)$$

where $x_0 \equiv k_0\xi_n$; $j_0$, $j_1$, $\eta_0$ and $\eta_1$ are spherical Bessel and Neumann functions of orders zero and one; $P_0(y)$ and $P_1(y)$ are the Legendre polynomials. The phases of the s and p waves for the neutron energy corresponding to $k_0$ are given by the following expressions [6]:

$$\exp(2i\delta_{0,1}(k_0)) = \exp(2i\delta_{0,1}^{pot}(k_0))\left(1 - \frac{i\Gamma_{n0,1}}{\varepsilon_p - E_{0,1} + \frac{i}{2}\Gamma_{n0,1}}\right), \quad (13)$$

where $\Gamma_{n0,1}$ are the elastic s- and p-resonance widths for the neutrons, $E_{0,1}$ are the positions of the s and p resonances, $\varepsilon_p$ is the energy of the incoming neutron, and $\delta_{0,1}^{pot}$ are the potential scattering phases of the s and p waves at the neutron energy corresponding to $k_0$. The wave function $\varphi_{fin}$ of the scattered neutron will also have the form of a superposition of s and p waves. The expression for $\varphi_{fin}$ has the form of (12) with $x_0$ replaced by $x_1 \equiv k_1\xi_n$, where $k_1^2 = 2M/(\varepsilon_p \pm \hbar\omega)/\hbar$ and the scattering phases $\delta_0$ and $\delta_1$ evaluated at the neutron energy corresponding to $k_1$.

We shall compute the radial matrix element of the transition operator (10) between the wave functions $\varphi_{init}$ and $\varphi_{fin}$ in the square well model. Using the well-known quantum mechanical formula

$$(\ddot{\xi}_n)_{12} = (E_1 - E_2)(\xi_n)/\hbar^2, \quad (14)$$

as well as the relation

$$M(\ddot{\xi}_n)_{\lambda n} = -(\nabla_n U)_{\lambda n}, \quad (15)$$

which is valid for the Hamiltonian (4), we obtain for the differential cross section for neutron scattering in a laser-radiation field the expression

$$\frac{d\sigma_{1n}}{d\Omega_{kt}} = \left(\frac{e_t E_0}{\hbar\omega}\right)^2 \left(\frac{U_0}{\hbar\omega}\right)^2 \Delta^2 \frac{1}{k_0 k_1}$$
$$\times [|A|^2 \cos^2\theta + |B|^2 (\cos\theta_0\cos\theta_1 + \sin\theta_0\sin\theta_1\cos(\varphi_1 - \varphi_0))^2 \quad (16)$$
$$-2(\operatorname{Re}A\operatorname{Re}B + \operatorname{Im}A\operatorname{Im}B)(\cos\theta_0\cos\theta_1 + \sin\theta_0\sin\theta_1\cos(\varphi_1 - \varphi_0))\cos\theta_0],$$

where

$$A = \frac{k_1}{k_0}(R + a_0(k_1))\frac{\delta_1(k_0)}{k_0 R}, \quad B = (R + a_0(k_0))\frac{\delta_1(k_1)}{k_1 R}, \tag{17}$$

$U_0$ is the depth of the well, $\theta_1$ is the angle between $\mathbf{k}_0$ and $\mathbf{k}_1$, and $a_0(k_0) = \delta_0(k_0)/k_0$, $a_0(k_1) = \delta_0(k_1)/k_1$ are the s-wave neutron scattering lengths at $\varepsilon_p$ and $\varepsilon_p \pm \hbar\omega$ respectively.

Integrating the expression (16) over the scattering angle, we obtain the total cross section for inelastic neutron scattering in a laser-radiation field:

$$\sigma_{\lambda n} = \left(\frac{e_t E_0}{\hbar\omega}\right)^2 \left(\frac{U_0}{\hbar\omega}\right)^2 |\Delta|^2 \left[|A|^2 \cos^2\theta_0 + \frac{1}{3}|B|^2\right] \frac{4\pi}{k_0 k_1}. \tag{18}$$

Let us consider in detail the two simple cases in which one of the terms in the expression (18) is greater than the other. Let the initial neutron energy $\varepsilon_p$ be such that it practically coincides with the s-resonance energy $E_0$ of the compound nucleus, while the scattered-neutron energy lies in the p-resonance region ($\varepsilon_p \pm \hbar\omega \approx E_1$). Then, since the resonance scattering phase is usually much greater than the potential scattering phase, $|B|^2 >> |A|^2$ and

$$\sigma_{\lambda n}^{(1)} \approx \left(\frac{e_t E_0}{\hbar\omega}\right)^2 \left(\frac{U_0}{\hbar\omega}\right)^2 |\Delta|^2 |B|^2 \cos^2\theta_0 \frac{4\pi}{3 k_0 k_1},$$

$$|B|^2 \approx \frac{\Gamma_{n0}^2(k_1)/k_1^2}{4(\varepsilon_p - E_0)^2 + \Gamma_0^2(k_0)} \frac{\Gamma_{n1}^2(k_1)/(k_1 R)^2}{4(\varepsilon_p \pm \hbar\omega - E_1)^2 + \Gamma_1^2(k_1)}. \tag{19}$$

In the second case $\varepsilon_p \approx E_1$ and $\varepsilon_p \pm \hbar\omega \approx E_1$, i.e., the initial neutron energy is close to the p resonance, while the scattered- neutron energy is close to the s resonance. Then the total scattering cross section has the form

$$\sigma_{\lambda n}^{(2)} \approx \left(\frac{e_t E_0}{\hbar\omega}\right)^2 \left(\frac{U_0}{\hbar\omega}\right)^2 |\Delta|^2 |A|^2 \cos^2\theta_0 \frac{4\pi}{k_0 k_1},$$

$$|A|^2 \approx \frac{k_1^2}{k_0^2} \frac{\Gamma_{n0}^2(k_1)/k_1^2}{4(\varepsilon_p \pm \hbar\omega - E_0)^2 + \Gamma_0^2(k_1)} \frac{\Gamma_{n1}^2(k_0)/(k_0 R)^2}{4(\varepsilon_p - E_1)^2 + \Gamma_1^2(k_0)}. \tag{20}$$

Thus, as follows from a comparison of the formulas (19) and (20), in the second case (transition from the p into the s state) the total cross section depends on the angle $\theta_0$ between the vectors $\mathbf{k}_0$ and $\mathbf{E}_0$ and, when $\theta_0 = \pi/2$ the total inelastic neutron scattering cross section tends to zero. At the same time, in the first case (transition from the **s** into the p state) the total cross section does not depend on this angle at all. Let us estimate the ratio of the total cross sections at exact resonance, when in the first case the initial neutron energy coincides with the s-resonance energy and the scattered neutron energy coincides with the location of the p resonance (let, for definiteness, the transition from the s into the p state be accompanied by the emission of a photon). Then the transition from the p into the s state is accompanied by the absorption of a photon:

$$\frac{\sigma_{\lambda n}^{(2)}}{\sigma_{\lambda n}^{(1)}} \sim 3\cos^2\theta_0 \frac{\varepsilon_p + \hbar\omega}{\varepsilon_p} > 1, \qquad (21)$$

if $\cos^2\theta_0 \sim 1$.

## 6. DISCUSSION OF THE RESULTS. CONCLUSION

Above we determined the neutron-scattering and reaction cross sections in the field of an electromagnetic wave whose frequency is equal to the separation of the compound nucleus levels with opposite parities. But a number of conditions must be fulfilled to be able to observe the effect of laser radiation on the cross section for interaction of a neutron with nuclei.

1. The neutron must be scattered by the nucleus of an ion. For example, atoms of the rare-earth elements and the actinides, when implanted in a dielectric, exist in an ionic state with charge $Z - Z' \sim 3$; therefore, the effective charge of the neutron + nucleus system is, according to (18), reduced by a factor of 20-30 as a result of the effect of the electron shell. But for the ion to be regarded as isolated, the condition $\omega \gg \omega_{opt}$, where $\omega_{opt}$ is the characteristic frequency of the optical phonons, must be fulfilled; otherwise, according to the conclusion drawn in Sec. 2, the effective charge can drop down to $10^{-4}$.

Let us, moreover, note that the proposed method of computing the cross sections is valid if the condition $\hbar\omega \gg \Gamma_0$ is also fulfilled.

2. The scattering, accompanied by the absorption or emission of a quantum of the electromagnetic field, of an s wave neutron near an isolated p resonance is highly improbable, since in this case the s-wave neutron undergoes only potential scattering, and does not form a compound nucleus in the potential-well model.

3. Estimates with the aid of the formulas (19) and (20) show that the cross section for s-wave neutron scattering accompanied by a transition of the neutron into the p state is of the same order of magnitude as the corresponding cross section for the case in which the p-wave neutron goes over into the s state. But the total cross section for capture in the case of the s resonance of the compound nucleus is, as a rule, greater than the corresponding cross section in the case of the neighboring p level. Therefore, it is more expedient to have the p level of the compound nucleus in the initial state, since we can use a thicker target in that case. Let us also emphasize that the transition from the s(p) level of the compound nucleus into another state, which may be forbidden or of low probability in the absence of an electromagnetic field (the partial (n, $\gamma$) transition, for example), can become allowed to the extent of the mixing of the compound-nucleus levels after the field has been switched on.

The case in which the external electromagnetic wave is at resonance with some electron transition in the atom (ion) is of interest. Such a resonance in the Rabi regime ($\gamma_{el} << \Delta\omega$, the electromagnetic field width) is considered in [2], where it is shown that in this case the high-frequency electromagnetic field inside the electron shell can be an appreciably amplified field. Resonance in the case when the Rabi conditions are not fulfilled requires a more detailed analysis.